\documentclass{article}
\usepackage{graphicx}
\usepackage{cite}

\begin{document}

\title{Scheduling Wireless Links by Graph Multicoloring in the Physical
Interference Model}

\author{Fabio~R.~J.~Vieira\thanks{Corresponding author
(fabiorjvieira@gmail.com).}\\
Jos\'e~F.~de~Rezende\\
Valmir~C.~Barbosa\\
\\
Programa de Engenharia de Sistemas e Computa\c c\~ao, COPPE\\
Universidade Federal do Rio de Janeiro\\
Caixa Postal 68511\\
21941-972 Rio de Janeiro - RJ, Brazil}

\date{}

\maketitle

\begin{abstract}
Scheduling wireless links for simultaneous activation in such a way that all
transmissions are successfully decoded at the receivers and moreover network
capacity is maximized is a computationally hard problem. Usually it is tackled
by heuristics whose output is a sequence of time slots in which every link
appears in exactly one time slot. Such approaches can be interpreted as the
coloring of a graph's vertices so that every vertex gets exactly one color. Here
we introduce a new approach that can be viewed as assigning multiple colors to
each vertex, so that, in the resulting schedule, every link may appear more than
once (though the same number of times for all links). We report on extensive
computational experiments, under the physical interference model, revealing
substantial gains for a variety of randomly generated networks.

\end{abstract}

\newpage
\section{Introduction}\label{intr}

Let $L$ be a set of wireless links, each link $i\in L$ characterized by a sender
node $s_i$ and a receiver node $r_i$. Depending on the spatial disposition of
such nodes, activating more than one link simultaneously creates interference
that may hamper the receivers' ability to decode what they receive. In the
physical interference model \cite{Gupta2000}, the chief quantity governing
receiver $r_i$'s ability to decode what it receives from $s_i$ when all links of
a set $S$ containing link $i$ are active is the signal-to-interference-and-noise
ratio (SINR), given by
\begin{equation}
\mathrm{SINR}(r_i,S)=
\frac{P/d^\alpha_{s_ir_i}}
{N+\sum_{j\in S\setminus\{i\}}P/d^\alpha_{s_jr_i}},
\end{equation}
where $P$ is a sender's transmission power (assumed the same for all senders),
$N$ is the noise floor, $d_{ab}$ is the Euclidean distance between nodes $a$ and
$b$, and $\alpha>2$ determines the law of power decay with Euclidean distance.
We say that a nonempty subset $S$ of $L$ is \emph{feasible} if no two of its
members share a node (in case $\vert S\vert>1$) and moreover
$\mathrm{SINR}(r_i,S)\ge\beta$ for all $i\in S$, where $\beta$ is a parameter
related to a receiver's decoding capabilities (assumed the same for all
receivers) and is chosen so that $\beta>1$.

Several strategies have been devised to maximize network capacity, either
through the self-contained scheduling of the links in $L$ for activation
\cite{Jain2003,
Brar2006,
Wang2006,
Moscibroda2007,
Chafekar2008,
Hua2008,
Goussevskaia2009,
Santi2009,
Boyaci2010,
Yang2010,
Augusto2011,
Leconte2011,
Shi2011,
Goussevskaia2014} or by combining link scheduling with other techniques
\cite{Cruz2003,
Alicherry2005,
Wang2008,
Capone2010,
Kesselheim2011,
Rubin2012,
Vieira2012}. All these strategies revolve around formulations as NP-hard
optimization problems, so all rely on some form of heuristic procedure drawing
inspiration from various sources, some merely intuitive, others more formally
grounded on graph-theoretic notions. Often the problem is formulated in a
spatial time-division multiple access (STDMA) framework, that is, assuming
essentially that time is divided into time slots, each one accommodating a
certain number of simultaneous link activations. In this case, the problem is to
find $T$ feasible subsets of $L$, here denoted by $S_1,S_2,\ldots,S_T$,
minimizing $T$ while ensuring that every link appears in exactly one of the $T$
subsets.

There is a sense in which this formulation can be interpreted in the context of
coloring a graph's vertices. Specifically, if we regard the links to be
scheduled as vertices in a graph, and furthermore say that no two vertices of a
group are neighbors of each other if the corresponding set of links is feasible,
then the schedule given by the sequence $S_1,S_2,\ldots,S_T$ of feasible link
sets establishes a coloring of the graph's vertices with $T$ colors in which all
vertices in $S_k$ get color $k$. This interpretation suggests a generalization
of the above formulation that requires every link to appear not in exactly one
of the $T$ subsets but in any number $q$ of subsets, provided this number is the
same for all links. In this generalized formulation, the goal is no longer to
minimize $T$, but rather to find the values of $T$ and $q$ that minimize the
ratio $T/q$. Returning to the vertex-coloring interpretation, now a vertex
receives $q$ (out of $T$) distinct colors, each relative to the time slot in
which the corresponding link is scheduled to be activated (i.e., $q$ of the
subsets $S_1,S_2,\ldots,S_T$).

The potential advantages of this multicoloring-based formulation are
tantalizing. If the original formulation leads to a number $T$ of slots while
the new one leads to $T'>T$ slots for some $q>1$, the latter schedule is
preferable to the former, even though it requires more time slots, provided only
that $T'/q<T$ (or $qT>T'$). To see that this is so, first note that the longer
schedule promotes an overall number of link activations given by $q\vert L\vert$
in $T'$ time slots. In order for the shorter schedule to achieve this same
number of activations, it would have to be repeated $q$ times in a row, taking
up $qT>T'$ time slots.

The possibility of multicoloring-based link scheduling in the physical
interference model seems to have been overlooked so far, despite the recent
demonstration of its success in the protocol-based interference model
\cite{Vieira2012}. Here we introduce a heuristic framework to obtain
multicoloring-based schedules from the single-color schedules produced by any
rank-based heuristic (i.e., one that decides the time slot in which to activate
a given link based on how it ranks relative to the others with respect to some
criterion). We use two iconic single-color heuristics (GreedyPhysical
\cite{Brar2006}, for its simplicity, and ApproxLogN
\cite{Goussevskaia2009,Goussevskaia2014}, for its role in establishing new
bounds on network capacity), as well as a third one that we introduce in
response to improvement opportunities that we perceived in the former two.
Incidentally, the latter heuristic, called MaxCRank, is found to perform best
both as a stand-alone, single-color strategy and as a base for the multicoloring
scheme.

\section{Single-color schedules}\label{1color}

Rank-based heuristics for single-color scheduling are usually monotonic, in the
sense that first $S_1$ is determined, then $S_2$ out of the set $R$ of links
that remain to be scheduled, then $S_3$ out of a smaller $R$, and so on, until
$R$ becomes empty. Choosing a link to add to the current $S_k$ depends on the
feasibility of the resulting set and also on a ranking criterion that is
specific to each heuristic. The ranking criterion establishes the order in which
the links in $R$ are to be considered for inclusion in $S_k$.

The following is the general outline of such a heuristic.
\begin{enumerate}
\item Let $k:=1$, $S_k:=\emptyset$, and $R:=L$. Order $R$ according to the
ranking criterion.
\item If a link $i\in R$ exists such that $S_k\cup\{i\}$ is feasible, then move
the top-ranking such $i$ from $R$ to $S_k$ and go to Step~3. If none exists,
then let $k:=k+1$, $S_k:=\emptyset$, and go to Step~2.
\item If $R\neq\emptyset$, then reorder $R$ according to the ranking criterion
and go to Step~2.
\item Let $T:=k$ and output $S_1,S_2,\ldots,S_T$.
\end{enumerate}
Steps 1--4 amount to scanning the set $R$ of unscheduled links and moving to the
current $S_k$ (in Step~2) the top-ranking link $i\in R$ whose inclusion in $S_k$
preserves feasibility. Whenever such a move does occur, an opportunity is
presented for $R$ to be reordered (in Step~3) according to the ranking
criterion.

It is easy to see that both GreedyPhysical and ApproxLogN can be cast in this
sequence of steps in a straightforward manner. The ranking criterion for
GreedyPhysical is nonincreasing and refers, for link $i$, to the number of links
in $L$ with which $i$ can never share a time slot; that is, links $j\in L$ such
that $\{i,j\}$ is infeasible. It is then an immutable ranking criterion and
consequently the reordering in Step~3 is moot. As for ApproxLogN, its ranking
criterion is nondecreasing, but now refers to the Euclidean distance between the
sender and the receiver in each link. This criterion, too, is fixed and as such
renders the reordering in Step~3 once again moot.\footnote{ApproxLogN replaces
the requirement of feasibility in Step~2 by conditions that are sufficient for
it to be satisfied. This is done to make sure that certain algorithmic
performance guarantees hold, but that is of no concern to us here.}

We now introduce a new heuristic that can also be viewed as instantiating
Steps~1--4, but with a ranking criterion that is both more stringent than the
two just described and also inherently dynamic, thus justifying the reordering
in Step~3. We call it MaxCRank to highlight its core principle, which is to
maximize the number of links in $R$ that still have a chance of joining the
current $S_k$ (i.e., remain ``Candidates'') once a decision is made on which one
of them, say $i$, is to be moved from $R$ to $S_k$. The corresponding ranking
criterion is nondecreasing and refers to the number of links
$j\in R\setminus\{i\}$ for which $S_k\cup\{i,j\}$ is infeasible.

\section{Multicoloring-based schedules}\label{+color}

The link sets $S_1,S_2,\ldots,S_T$ output by Steps~1--4 of Section~\ref{1color} 
promote a number of link activations given by $\vert L\vert$, one activation per
link. If this schedule were to be repeated $q$ times in a row for some $q>1$,
the total number of link activations would grow by a factor of $q$ and so would
the number of time slots used. That the same growth law should apply both to
how many links are activated and to how many time slots elapse indicates that
the most basic scheduling unit is $S_1,S_2,\ldots,S_T$ itself, not any number of
repetitions thereof.

However, activating the links in $S_1$ the second time around does not
necessarily have to be restricted to time slot $T+1$. Instead, it may be
possible to take advantage of some room left in previous time slots for at least
one of the links in $S_1$. With this type of precaution in mind, advancing link
activations in such a manner might result in a sequence of link sets
$S_1,S_2,\ldots,S_{T'}$ containing exactly $q$ activations of every link in $L$
for some $q>1$ but with $T<T'<qT$. Clearly, in this case the most basic
scheduling unit would be $S_1,S_2,\ldots,S_{T'}$, not $S_1,S_2,\ldots,S_T$ any
more. Not only this, but the new basic scheduling unit would be preferable to
the previous one, since a total of $q\vert L\vert$ link activations would be
attainable in fewer time slots ($T'$ rather than $qT$).

A heuristic to find the greatest $q>1$ for which $T'<qT$, if any exists, is
simply to wrap Steps~1--4 in an outer loop that iterates along with
$q=1,2,\ldots$ while preventing $S_k$ from being reset to $\emptyset$ any later
than the first time it is considered. At the end of each iteration, say the
$q$th, the value of $T'$ is updated (to the number of time slots elapsed since
the beginning) and the ratio $T'/q$ is computed. The iterations continue while
this ratio is strictly decreasing. At the end of the first iteration we get
$T'=T$, but successful further iterations will produce a sequence of strictly
decreasing $T'/q$ values. A new quantity of interest is then the gain $G$
incurred by the resulting heuristic, that is, the ratio of $T$ to the last
$T'/q$, hence $G=qT/T'$. The least possible value of $G$, of course, is $G=1$,
which corresponds to the case in which the iterations fail already for $q=2$.

\section{Experimental setup}\label{setup}

We give results for two families of randomly generated networks, henceforth
referred to as type-I and type-II networks. As will become apparent, type-I
networks are more realistic. We use type-II networks as well because they were
used in the performance evaluation of ApproxLogN
\cite{Goussevskaia2009,Goussevskaia2014} and thus provide a more direct basis
for comparison. A network's number of nodes is henceforth denoted by $n$.

\begin{figure}[p]
\centering
\scalebox{1.0}{\includegraphics{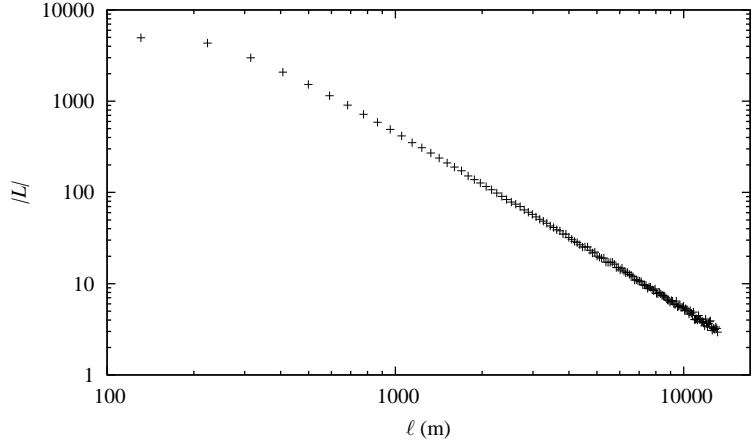}}
\caption{Average number of links in type-I networks as a function of the square
side $\ell$ for $n=100$. All data points are averages over $1\,000$ network
instances. Additional relevant parameters are $P=300$~mW,
$N\approx 8\times 10^{-14}$~W (for a bandwidth of $20$~MHz at room temperature),
$\alpha=4$, and $\beta=25$~dB.}
\label{nlinks}
\end{figure}

\begin{figure}[p]
\centering
\scalebox{1.0}{\includegraphics{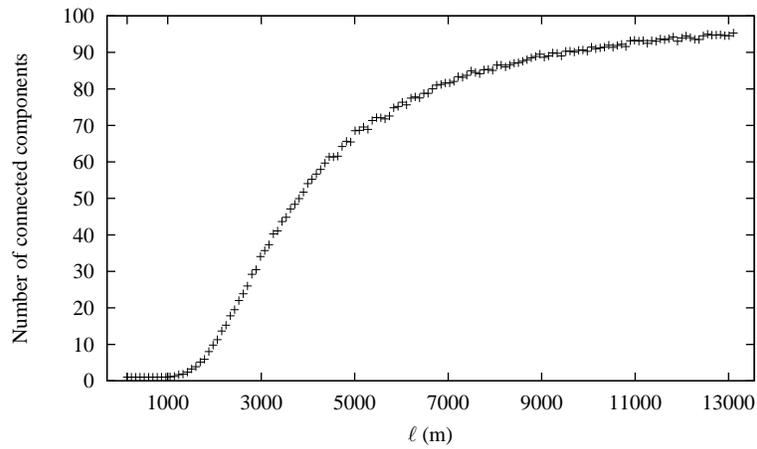}}
\caption{Average number of connected components in type-I networks as a function
of the square side $\ell$ for $n=100$. All data points are averages over
$1\,000$ network instances. Additional relevant parameters are $P=300$~mW,
$N\approx 8\times 10^{-14}$~W (for a bandwidth of $20$~MHz at room temperature),
$\alpha=4$, and $\beta=25$~dB.}
\label{ncc}
\end{figure}

A type-I network is generated by first placing all $n$ nodes inside a square of
side $\ell$ uniformly at random. A node's neighbors are then determined as a
function of the value of $d_{s_ir_i}$ for which
$\mathrm{SINR}(r_i,\{i\})=\beta$. Denoting such a distance by $\rho$ yields
$\rho=(P/\beta N)^{1/\alpha}$, so a node's neighbor set is the set of nodes to
which the Euclidean distance does not surpass $\rho$. Any two nodes that are
neighbors of each other become a link in $L$, sender and receiver being decided
uniformly at random (so that a node may, e.g., be the sender in a link and the
receiver in another). For fixed $n$, increasing $\ell$ causes the number of
links, $\vert L\vert$, to decrease precipitously, though in the heavy-tailed
manner of an approximate power law (Fig.~\ref{nlinks}). It also causes the
network's number of connected components to increase from about $1$ to nearly
$n$ (a component per node) through a sharp transition in between
(Fig.~\ref{ncc}).

In a type-II network, the number $n$ of nodes is necessarily even. Of these,
$n/2$ are senders and $n/2$ are receivers. A type-II network is generated by
first placing the receivers uniformly at random inside a square of side $\ell$
and then, for each receiver, placing the corresponding sender inside a circle
of radius $\rho$ centered at it, also uniformly at random. A type-II network has
$n/2$ links and connected components. Varying $\ell$ affects interference only.

\section{Results}\label{results}

We give results for all three single-color heuristics mentioned in
Section~\ref{1color}, namely GreedyPhysical, ApproxLogN, and MaxCRank, and also
for their multicoloring-based versions, obtained as explained in
Section~\ref{+color}. These results are given as $T/\vert L\vert$ in the former
case (the normalized schedule length, since $\vert L\vert$ is a clear upper
bound on $T$), and as the gain $G$ in the latter.

\begin{figure}[p]
\centering
\scalebox{1.0}{\includegraphics{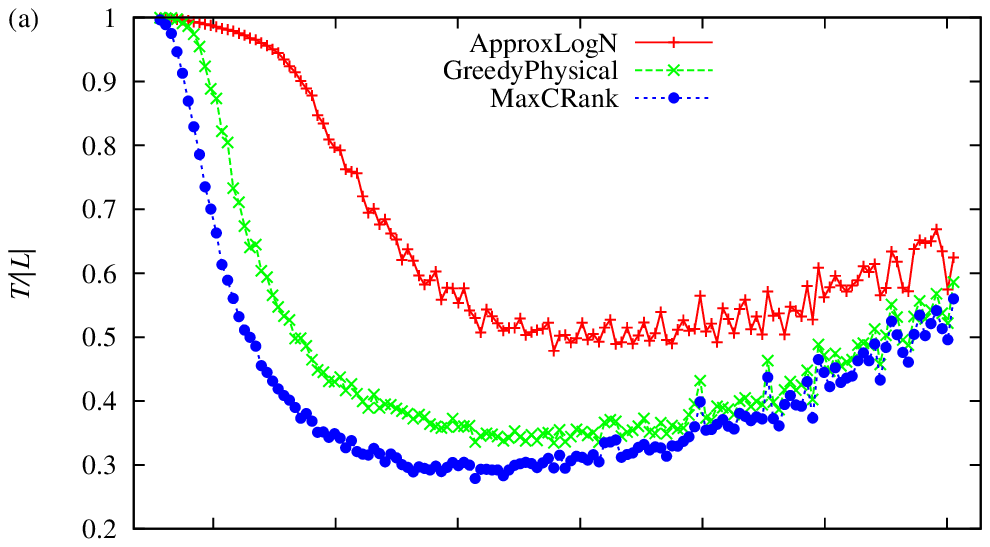}}
\\\vspace{0.05in}
\scalebox{1.0}{\includegraphics{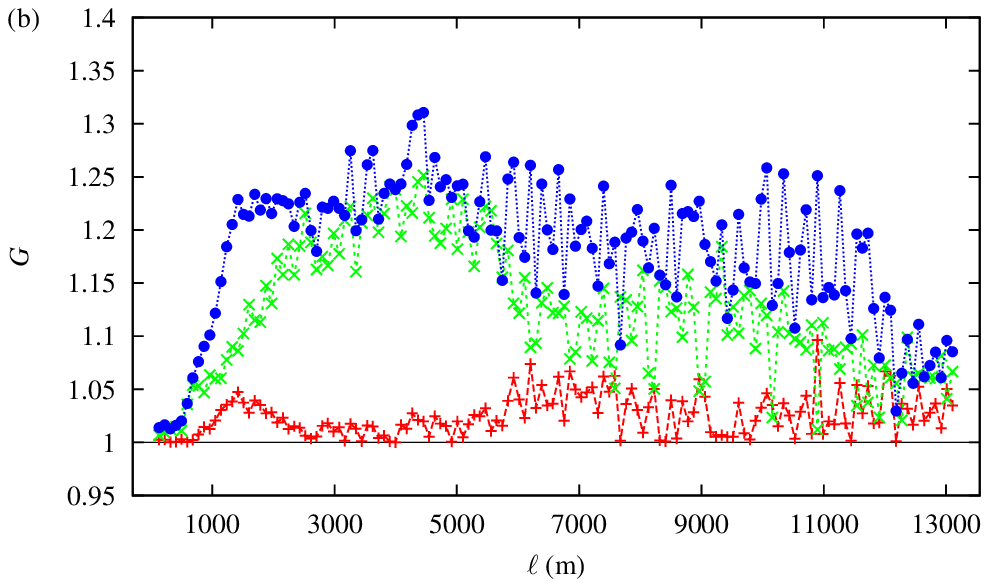}}
\caption{Performance of GreedyPhysical, ApproxLogN, and MaxCRank on type-I
networks. Data are given for the heuristics' single-color versions (a) and for
their multicoloring-based versions (b). All data points are averages over
$1\,000$ network instances. Confidence intervals are less than $1\%$ of the
mean at the $95\%$ level, so error bars are omitted. All networks have $n=100$
nodes. Additional relevant parameters are $P=300$~mW,
$N\approx 8\times 10^{-14}$~W (for a bandwidth of $20$~MHz at room
temperature), $\alpha=4$, and $\beta=25$~dB.}
\label{resultsI}
\end{figure}

The data in Fig.~\ref{resultsI} refer to type-I networks and as such are given
as a function of the square side $\ell$. The number of nodes is fixed throughout
(at $n=100$), so the networks get sparser (fewer links, more connected
components) as $\ell$ is increased. In the single-color cases (panel (a) of the
figure), all three heuristics start out with $T=\vert L\vert$ for the very dense
networks (very small $\ell$), but smaller densities quickly reduce interference
so that $T$ falls significantly below $\vert L\vert$. MaxCRank is the best
performer throughout, followed by GreedyPhysical and ApproxLogN. As for the
heuristics' multicoloring-based versions (panel (b)), there is practically no
gain for the densest networks, but again this is reversed as interference abates
with increasing $\ell$. MaxCRank is still the top performer and ApproxLogN the
bottom one (in fact, the only of the three heuristics for which $G=1$ is
sometimes attained).

\begin{figure}[p]
\centering
\scalebox{1.0}{\includegraphics{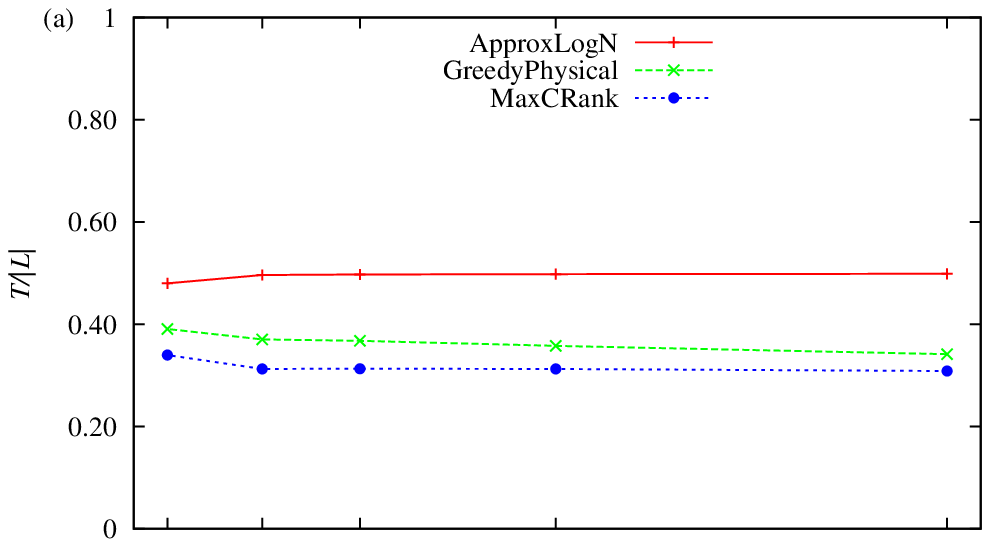}}
\\\vspace{0.05in}
\scalebox{1.0}{\includegraphics{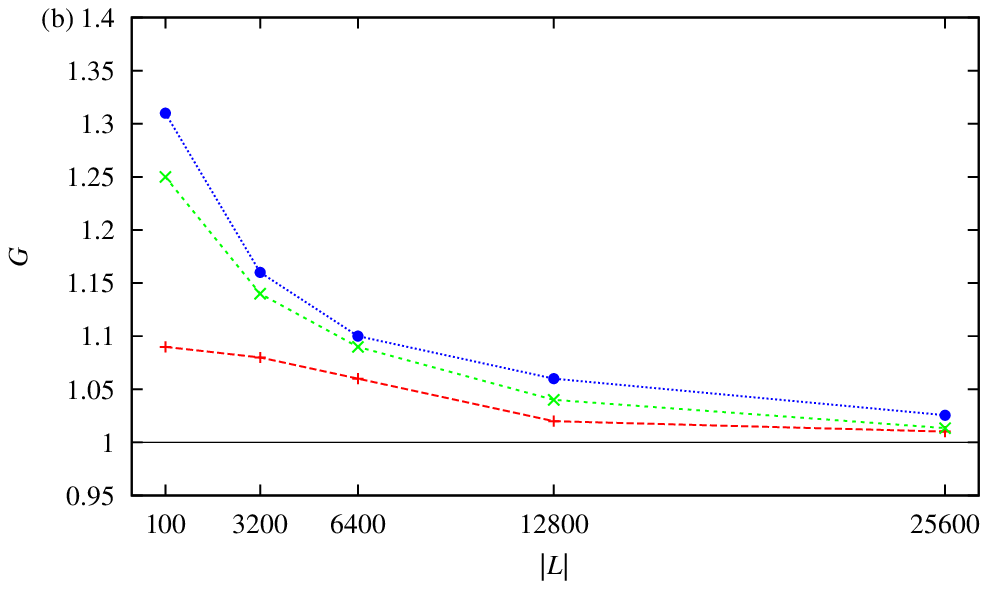}}
\caption{Performance of GreedyPhysical, ApproxLogN, and MaxCRank on type-II
networks. Data are given for the heuristics' single-color versions (a) and for
their multicoloring-based versions (b). All data points are averages over
$1\,000$ network instances. Confidence intervals are less than $1\%$ of the mean
at the $95\%$ level, so error bars are omitted. All networks have
$\ell=1\,000$~m. Additional relevant parameters are $P=300$~mW,
$N\approx 8\times 10^{-14}$~W (for a bandwidth of $20$ MHz at room temperature),
$\alpha=4$, and $\beta=25$~dB.}
\label{resultsII}
\end{figure}

The results for type-II networks, given in Fig.~\ref{resultsII}, are presented
as a function of $\vert L\vert=n/2$, the number of links. Because $\ell$ is
fixed throughout (at $\ell=1\,000$~m), increasing $\vert L\vert$ causes the
impact of accumulated interference to be felt more severely. One consequence of
this is that, for the single-color heuristics (panel (a) of the figure), $T$
increases almost linearly with $\vert L\vert$. Another consequence, now related
to the multicoloring-based versions of the heuristics (panel (b)), is that gains
above $1$ are increasingly hard to come by as $\vert L\vert$ is increased.
MaxCRank continues to be the top performer in all cases, followed by
GreedyPhysical, then by ApproxLogN.

\section{Discussion}\label{disc}

Although it may at first seem striking that ApproxLogN has performed so poorly
across most of our experiments, it should be kept in mind that this heuristic,
in all likelihood, was never meant as a serious contender for single-color link
scheduling. In fact, and as noted in Section~\ref{1color}, ApproxLogN approaches
the checking of feasibility rather indirectly, verifying sufficient conditions
for feasibility to hold instead of the property itself. This is bound to prevent
ApproxLogN from scheduling links for activation when they could be scheduled.
What must be kept in mind, then, is that the use of such indirect conditions has
led to important performance and capacity bounds. ApproxLogN, therefore, remains
an important contribution despite its performance in more practical settings.

What really is striking in our results, though, is the appearance of
greater-than-$1$ gains practically across the board, particularly for
MaxCRank or GreedyPhysical as the base, single-color heuristic. Link schedules,
once determined, are meant to be used repetitively, so every link is already
meant to be scheduled for activation over and over again, indefinitely.
Conceptually, what our multicoloring-based wrapping of single-color heuristics
tries to do is to intertwine some number of repetitions of a single-color
schedule, taking up fewer time slots than the straightforward juxtaposition of
the same number of repetitions of that schedule. By doing so, more link
activations can be packed together in earlier time slots. As a consequence, the
basic schedule to be used for indefinite repetition is now one that leads to
higher network capacity and possibly higher throughput.

As we mentioned earlier, multicoloring-based link scheduling of the sort we have
demonstrated has roots in the multicoloring of a graph's vertices (as well as
edges, in many cases). As such, a rich body of material, relating both to
computational-complexity difficulties and to workarounds in important cases, is
available. Further developments should draw on such knowledge, aiming to obtain
more principled, and perhaps even better performing, heuristics.

\subsection*{Acknowledgments}

The authors acknowledge partial support from CNPq, CAPES, and a FAPERJ BBP
grant.

\bibliography{mcsched}

\begin{thebibliography}{10}

\bibitem{Gupta2000}
P.~Gupta and P.~R. Kumar.
\newblock The capacity of wireless networks.
\newblock {\em IEEE T. Inform. Theory}, 46:388--404, 2000.

\bibitem{Jain2003}
K.~Jain, J.~Padhye, V.~N. Padmanabhan, and L.~Qiu.
\newblock Impact of interference on multi-hop wireless network performance.
\newblock In {\em Proc. MobiCom}, pages 66--80, 2003.

\bibitem{Brar2006}
G.~Brar, D.~M. Blough, and P.~Santi.
\newblock Computationally efficient scheduling with the physical interference
  model for throughput improvement in wireless mesh networks.
\newblock In {\em Proc. MobiCom}, pages 2--13, 2006.

\bibitem{Wang2006}
W.~Wang, Y.~Wang, X.-Y. Li, W.-Z. Song, and O.~Frieder.
\newblock Efficient interference-aware {TDMA} link scheduling for static
  wireless networks.
\newblock In {\em Proc. MobiCom}, pages 262--273, 2006.

\bibitem{Moscibroda2007}
T.~Moscibroda, Y.~A. Oswald, and R.~Wattenhofer.
\newblock How optimal are wireless scheduling protocols?
\newblock In {\em Proc. INFOCOM}, pages 1433--1441, 2007.

\bibitem{Chafekar2008}
D.~Chafekar, V.~S.~A. Kumart, M.~V. Marathe, S.~Parthasarathy, and
  A.~Srinivasan.
\newblock Approximation algorithms for computing capacity of wireless networks
  with {SINR} constraints.
\newblock In {\em Proc. INFOCOM}, pages 1166--1174, 2008.

\bibitem{Hua2008}
Q.-S. Hua and F.~C.~M. Lau.
\newblock Exact and approximate link scheduling algorithms under the physical
  interference model.
\newblock In {\em Proc. DIALM-FOMC}, pages 45--54, 2008.

\bibitem{Goussevskaia2009}
O.~Goussevskaia, R.~Wattenhofer, M.~M. Halldorsson, and E.~Welzl.
\newblock Capacity of arbitrary wireless networks.
\newblock In {\em Proc. INFOCOM}, pages 1872--1880, 2009.

\bibitem{Santi2009}
P.~Santi, R.~Maheshwari, G.~Resta, S.~Das, and D.~M. Blough.
\newblock Wireless link scheduling under a graded {SINR} interference model.
\newblock In {\em Proc. FOWANC}, pages 3--12, 2009.

\bibitem{Boyaci2010}
C.~Boyaci, B.~Li, and Y.~Xia.
\newblock An investigation on the nature of wireless scheduling.
\newblock In {\em Proc. INFOCOM}, pages 1--9, 2010.

\bibitem{Yang2010}
D.~Yang, X.~Fang, G.~Xue, A.~Irani, and S.~Misra.
\newblock Simple and effective scheduling in wireless networks under the
  physical interference model.
\newblock In {\em Proc. GLOBECOM}, pages 1--5, 2010.

\bibitem{Augusto2011}
C.~H.~P. Augusto, C.~B. Carvalho, M.~W.~R. da~Silva, and J.~F. {de Rezende}.
\newblock {REUSE}: a combined routing and link scheduling mechanism for
  wireless mesh networks.
\newblock {\em Comput. Commun.}, 34:2207--2216, 2011.

\bibitem{Leconte2011}
M.~Leconte, J.~Ni, and R.~Srikant.
\newblock Improved bounds on the throughput efficiency of greedy maximal
  scheduling in wireless networks.
\newblock {\em IEEE/ACM T. Network.}, 19:709--720, 2011.

\bibitem{Shi2011}
Y.~Shi, Y.~T. Hou, S.~Kompella, and H.~D. Sherali.
\newblock Maximizing capacity in multihop cognitive radio networks under the
  {SINR} model.
\newblock {\em IEEE T. Mobile Comput.}, 10:954--967, 2011.

\bibitem{Goussevskaia2014}
O.~Goussevskaia, M.~M. Halldorsson, and R.~Wattenhofer.
\newblock Algorithms for wireless capacity.
\newblock {\em IEEE/ACM T. Network.}, 22:745--755, 2014.

\bibitem{Cruz2003}
R.~L. Cruz and A.~V. Santhanam.
\newblock Optimal routing, link scheduling and power control in multihop
  wireless networks.
\newblock In {\em Proc. INFOCOM}, pages 702--711, 2003.

\bibitem{Alicherry2005}
M.~Alicherry, R.~Bhatia, and L.~Li.
\newblock Joint channel assignment and routing for throughput optimization in
  multi-radio wireless mesh networks.
\newblock In {\em Proc. MobiCom}, pages 58--72, 2005.

\bibitem{Wang2008}
J.~Wang, P.~Du, W.~Jia, L.~Huang, and H.~Li.
\newblock Joint bandwidth allocation, element assignment and scheduling for
  wireless mesh networks with {MIMO} links.
\newblock {\em Comput. Commun.}, 31:1372--1384, 2008.

\bibitem{Capone2010}
A.~Capone, G.~Carello, I.~Filippini, S.~Gualandi, and F.~Malucelli.
\newblock Routing, scheduling and channel assignment in wireless mesh networks:
  optimization models and algorithms.
\newblock {\em Ad Hoc Netw.}, 8:545--563, 2010.

\bibitem{Kesselheim2011}
T.~Kesselheim.
\newblock A constant-factor approximation for wireless capacity maximization
  with power control in the {SINR} model.
\newblock In {\em Proc. SODA}, pages 1549--1559, 2011.

\bibitem{Rubin2012}
I.~Rubin, C.-C. Tan, and R.~Cohen.
\newblock Joint scheduling and power control for multicasting in cellular
  wireless networks.
\newblock {\em EURASIP J. Wirel. Comm.}, 2012:250, 2012.

\bibitem{Vieira2012}
F.~R.~J. Vieira, J.~F. {de Rezende}, V.~C. Barbosa, and S.~Fdida.
\newblock Scheduling links for heavy traffic on interfering routes in wireless
  mesh networks.
\newblock {\em Comput. Netw.}, 56:1584--1598, 2012.

\end{thebibliography}
\bibliographystyle{unsrt}

\end{document}